\documentstyle[12pt]{article}
\newcommand{\beq}{\begin{equation}}
\newcommand{\eeq}{\end{equation}}
\newcommand{\beqa}{\begin{eqnarray}}
\newcommand{\eeqa}{\end{eqnarray}}
\newcommand{\ba}{\begin{array}}
\newcommand{\ea}{\end{array}}

\begin{document}

\begin{flushright}
Preprint CAMTP/99-6\\
October 1999\\
\end{flushright}

\parindent 0pt
\vskip 0.5 truecm
\begin{center}
\Large
{\bf On 1D Schr\"odinger problems allowing polynomial solutions}\\
\vspace{0.25in}
\normalsize
George Krylov$^{\dag \, \ddag}$ and Marko Robnik$^{\dag}$
\footnote{e--mails:
george.krylov@uni-mb.si, robnik@uni-mb.si}\\

\vspace{0.3in}
\dag Center for Applied Mathematics and Theoretical Physics,\\
University of Maribor, Krekova 2, SI-2000 Maribor, Slovenia\\
\ddag Department of Physics, Belarusian State University,\\
Fr. Skariny av. 4, 220050 Minsk, Belarus\\

\end{center}
\parskip 10pt

\vspace{0.3in}

\normalsize
\noindent
{\bf Abstract.}
 {We discuss the explicit construction of the Schr\"odinger
 equations admitting representation through some family of general
 polynomials. Almost all solvable quantum potentials
 are shown to be generated by this approach. Some generalization
 has been performed also in higher dimensional problems. }

\vspace{0.6in}

Published in {\bf Journal of Physics A: Mathematical and General}\\
Vol. 33 (2000) 1233-1245
\normalsize
\vspace{0.1in}

\newpage

\section{Introduction}

In the past few years supersymmetric quantum mechanics 
based on shape invariance and
intertwining concepts has manifested a significant progress 
\cite{Fernandez-et-al95}-\cite{Das-Chakrabarti99}. 
Its technique started to
influence not only the traditional branches of physics 
such as atomic, nuclear and
high energy physics, which originally stimulated its 
emergence \cite{Witten81}, 
but also the classical areas of
mathematical physics and the theory of differential equations. 
Recently, in 
\cite{Jafarizadeh-Fakhri97}-\cite{Jafarizadeh-Fakhri98}
the detailed investigation of factorization 
technique has been performed for one specific form
of second order differential equations (SODE) with polynomial
coefficients, admitting polynomial solutions based on known Rodrigue's
formula \cite{Morse}. We choose  similar initial arguments to 
construct explicitly a wide class of  QM potentials 
for 1D Schr\"odinger equations admitting
(after separation of asymptotic behaviour of the 
wave function) polynomial solutions.

Our consideration and analysis in sections 2 and 3 shares 
a common subject with the known Natanzon
papers \cite{Natanzon71},\cite{Natanzon79} but differs from them
by being more general because of the fact that we work with Papperitz 
rather than hypergeometric equation with arbitrary positions of
the singular points (in complex domain). Moreover, in addition to
transformations described in \cite{Natanzon71},\cite{Natanzon79}
where the spectral parameters are preserved, we include into 
consideration the cases when we change the roles of spectral parameters
in original polynomial system and in the Schr\"odinger equation
(generalizing in this way the known consideration of the Coulomb
case, see the discussion in the text).

One more thing we would like to mention is that Turbiner's
approach \cite{Turbiner92} to generalized Bocher problem
is very near to ours for exactly solvable case (though in
\cite{Turbiner92} the possibility of the transformation to
appropriate Sturm-Liouville problem is merely mentioned rather than
investigated in full details). But for quasi-solvable cases
our approach differs from those in \cite{Turbiner92} because we
do not use the factorization ideas and investigate symmetry
properties but concentrate on the connection of the appropriate
polynomials with the corresponding family of 
Schr\"odinger equations. It allows us to explicitly define
the additional relations for the operator to be zero-grading
(in terms of  \cite{Turbiner92}) and gives the way of regular
construction of quasi-solvable potentials (all
inside some definite family) with an arbitrary chosen 
number of algebraically constructed eigenstates.
 
The last remark we should make is that we do not intend to
perform the comparison of the proposed approach with all other
known methods of the construction of exactly solvable quantum
potentials as this should be definitely the topic of a review paper
rather than original research paper. But, in our opinion, the proposed
approach will definitely influence the reviewing of the
results obtained in other ways (especially for quasi-solvable
cases) and further establishment of
its inherent relations with other methods.

The paper is organized as follows. In the Section 2
 we start with the special case leading to
polynomial solutions, namely the polynomial family introduced in 
\cite{Jafarizadeh-Fakhri97}-\cite{Jafarizadeh-Fakhri98}. Section 3
is devoted to the explicit construction of 
the Schr\"odinger equation corresponding to that
polynomial family, and to the presentation of some classification scheme as
well as
the discussion of its relation to the known solvable cases. 
In Section 3 we represent the  generalization of
 the proposed scheme in some directions,
namely the application to the partial differential equations and
 to higher order polynomial coefficients, and we  
demonstrate one  non-trivial irreducible example and make some concluding
remarks on the applicability of the considered approach.

\section{One Construction of Polynomial Solutions for SODE}
Practically all solvable 1D quantum problems correspond to those 
which can be transformed to the
equation of hypergeometric type, and that gives eigenvalues
 of bound states by the requirement
of finite hypergeometric series, thus being  a
polynomial of a given order 
\cite{Morse,Landau3}. 
We start the consideration from the case leading
 precisely to that known situation,
though from a little bit another point of view.
Recently \cite{Jafarizadeh-Fakhri97}-\cite{Jafarizadeh-Fakhri98} it
was shown that the eigenvalue problem for the operator of the form
\begin{eqnarray}  \label{eq:SA-LO1}
 \hat {\cal  L}= \frac{1}{W(x)}
\frac{d}{dx}\left(A(x)W(x)\frac{d}{dx}\right)
\end{eqnarray}
leads to  polynomial solutions with special
 requirements for the functions $A,W$.
Namely, if we 
choose $A(x)$ as a polynomial of at most a second order,
let us define,
\begin{equation}\label{eq:ax}
 A(x)=a_0+a_1 x+ a_2 x^2,
\end{equation}
 and $W(x)$ as a non-negative function
such that  $\frac{1}{W(x)}\frac{d}{dx}\left(A(x)W(x)\right)$ 
is at most a first order polynomial
\begin{equation}\label{eq:ax1}
 B(x)=b_0+b_1 x,
\end{equation}
then we can construct an orthogonal polynomial family
 being a solution for the eigenvalue problem, namely for
the operator $\hat {\cal L}$,
\begin{eqnarray}  \label{eq:SA-LO3}
  \frac{1}{W(x)}\frac{d}{dx}\left(A(x)W(x)
\frac{d\Phi_n(x)}{dx}\right) + \gamma_n \Phi_n(x)=0.
\end{eqnarray}
The polynomials given by the  classical Rodrigue's
formula \cite{Morse}
\begin{eqnarray}  \label{eq:rodrigues}
 \Phi_n(x)= \frac{a_n}{W(x)}\left(\frac{d}{dx}\right)^n
\left(A^n(x)W(x)\right)
\end{eqnarray}
are orthogonal with respect to the weight 
function W(x) on the interval $(a,b)$,
chosen such that the following conditions  hold 
\begin{equation}\label{boundary}
 A(a)W(a)=A(b)W(b)=0. 
\end{equation}
This can be shown by classical consideration 
as exposed e.g. in \cite{Morse}.
In this way
we can choose the interval as inside the roots 
of the polynomial $A(x)$,
 if the latter one possesses the real roots,
or the roots of $W(x)$, including infinity points,
 in the case of the functions tending to zero at infinity, or some
combination of both alternatives.
The eigenvalues $\gamma_n$ turn out to be given as 
\cite{Jafarizadeh-Fakhri97}-\cite{Jafarizadeh-Fakhri98}
\begin{eqnarray}\label{eq:gamma-n}
  \gamma_n=-n\left(\frac{(A(x)W(x))'}{W(x)}\right)'
 -\frac{n(n-1)}{2}A''(x), 
\end{eqnarray}
which in case (\ref{eq:ax}) and (\ref{eq:ax1}) is equal to expression (11).

At that point, having enough information,
 we can write and solve the  equation  for  $W(x)$, namely
\begin{eqnarray} \label{eq:W}
 \frac{d}{dx}\left(A(x)W(x)\right)-B(x)W(x)=0
\end{eqnarray}
As the equation is linear ODE of the first 
order its solution has the form,
explicitly
\begin{eqnarray} \label{eq:W1}
 W(x)=\frac{C}{A(x)}\exp \left\{\int 
\frac{B(x)dx}{A(x)} \right\}=
 \frac{C\exp \left\{\int \frac{b_0+b_1 x}{a_0+a_1 x+ a_2 x^2}dx 
\right\}}{a_0+a_1 x+ a_2 x^2},
\end{eqnarray}
and, of course, the integral we wrote is easy to calculate
\begin{eqnarray} \label{eq:int}
 \int \frac{b_0+b_1 x}{a_0+a_1 x+ a_2 x^2}dx=    {\frac{b_{1}
      \log (a_0 + a_1 x + a_2 x^2 )}{2
      a_{2}}}
-\\ \nonumber
{\frac{\left( -2a_{2}
          b_{0} + 
         a_{1}b_{1} \right) 
        \arctan ({\frac{a_{1} + 
            2a_{2}x}{{\sqrt{-{{
                 a_{1}}^2} + 
               4a_{0}a_{2}}
             }}})}{a_{2}
       {\sqrt{-{{a_{1}}^2} + 
           4a_{0}a_{2}}}}},
\end{eqnarray}
but for our purposes it is convenient to use 
it in the form we represent $W(x)$ in equation (\ref{eq:W1}).

It is evident from the direct substitution that in the case 
(\ref{eq:ax}) and (\ref{eq:ax1}) the
 eigenvalues $\gamma_n$ are given explicitly by
\begin{eqnarray}\label{eq:gamma}
\gamma_n=-n(b_1+a_2(n-1)),
\end{eqnarray}
while the equation for the polynomials becomes
\begin{eqnarray}\label{eq:Fi}
 (a_0+a_1 x+ a_2 x^2) \Phi_{n}''(x) +
(b_0+b_1 x)\Phi_{n}'(x) \;\; \\ \nonumber
-\; n(b_1+a_2(n-1))\Phi_{n}(x)=0
\end{eqnarray}
The equation (\ref{eq:Fi}) is SODE with 3 
singular points at the roots of $A(x)$ 
(we denote them as $x_1,x_2$) and infinity. 
It is easy to check that if the roots
of $A(x)$ are different the equation is of
 hypergeometric type (generally speaking
Papperitz equation \cite{Morse}), whilst it is  confluent
 hypergeometric when the roots 
are coincident.

The question naturally appearing is what sort 
of quantum mechanical problems could be associated with
the polynomial family we described. To answer it, 
there is a natural way, namely we can try to implement
adjusted pair of coordinate transformation and 
similarity transformation for the
equation (\ref{eq:Fi}) in such a way, as to obtain the
constant coefficient at second derivative and 
zero coefficient at first derivative. 
The resulting equation
will be of Schr\"odinger type. Let us perform this program.

\section{Transformation to the Schr\"odinger equation}
Let us forget for a while that $A(x)$ and $B(x)$ are 
polynomials. We have the equation
\begin{eqnarray}\label{eq:yx}
 A_2(x) y''(x) + A_1(x)y'(x)+ \varepsilon y(x)=0
\end{eqnarray}
with arbitrary coefficient functions $A_1(x),A_2(x)$.
First, we make the variable change 
\begin{eqnarray} \nonumber
 x=F(u) , \\\nonumber
\frac{d}{dx}=\frac{1}{F'(u)}\frac{d}{du}, \\ 
\frac{d^2}{dx^2}=\frac{1}{F'^2(u)}\frac{d^2}{du^2}-
\frac{F''(u)}{F'^3(u)}\frac{d}{du}, 
\end{eqnarray}
and choose the transformation in a form allowing 
to introduce some yet undefined
but prescribed function of new coordinate $\omega (u)$
 (which we could define later for the sake
of the most convenient choice) 
\begin{equation} \label{transf0}
\omega^2 (u) [F'(u)]^2=A_2(x)
\end{equation}
to get 
\begin{eqnarray}\label{eq:yu}
\omega^2(u) y''(u)+y'(u)\omega(u) \left( 
 \frac{2A_1(F(u))-A_2'(F(u))}{\sqrt{A_2(F(u))}}+\omega'(u)
\right) \\ \nonumber
+\gamma \,y(u) =0 
\end{eqnarray}
where prime means the differentiation with respect
 to the function's argument.
Now we implement the similarity transformation 
$Y(u)=\exp(\chi(u))y(u)$ 
and choose the function
$\chi(u)$ in such a way as to kill the term 
with the first derivative, so that we must have
\begin{eqnarray} \label{transf1}
\chi{\,'}(u)=\frac{1}{2\omega (u)} \left(  
  {\frac{{A_2}'(F(u))}
    {2\,{\sqrt{{A_2}(F(u))}}}}-
{\frac{{A_1}(F(u))}{\,{\sqrt{{A_2}(F(u))}}}}  
 - 
  \omega '(u) \right).
\end{eqnarray}
Then, the equation is transformed into 
\begin{eqnarray}\label{eq:Sch0}
\omega^2(u) Y''(u)+Y(u) \left(
\gamma  - {\frac{{{{A_1}(F(u))}^2}}{4\,{A_2}(F(u))}} 
 + 
  {\frac{{A_1}(F(u))\,{A_2}'(F(u))}
    {2\,{A_2}(F(u))}} 
\; \;\;\ \right. \\ \nonumber \left.
 - \; {\frac{3\,{{{A_2}'(F(u))}^2}}{16\,{A_2}(F(u))}} 
+ {\frac{{{\omega '(u)}^2}}{4}} + {\frac{{A_2}''(F(u))}{4}} 
- {\frac{{A_1}'(F(u))}{2}}
- {\frac{\omega (u)\,\omega ''(u)}{2}}  \right)
 \end{eqnarray}
which can be considered as
Schr\"odinger type equation if we manage to identify
 and to separate some
free  constant parameter in it at $Y(u)$ (playing 
the role of "energy"), 
after the division of  both terms by
$\omega^2(u)$. 
As we see in the trivial choice of
$\omega(u)=1$ we simply get the Schr\"odinger equation
 with the energy $\gamma$ and the potential
$V(u)$ given by
\begin{eqnarray}
V(u)=
  +{\frac{4\,{{A_1(F(u))}^2} - 
      8\,A_1(F(u))\,A_2'(F(u)) - 3\,{{A_2'(F(u))}^2}}
      {16\,A_2(F(u))}}  \\\nonumber
 -\; {\frac{A_1'(F(u))}{2}} - 
  {\frac{A_2''(F(u))}{4}}.
\end{eqnarray}
One known case when we have to choose $\omega(x) $ not 
equal to unity is the Coulomb potential
as we will see below.
Now we can use the fact that both $A_2(x),A_1(x)$ 
are  polynomials,
choosing them in accordance with (\ref{eq:ax},\ref{eq:ax1}) 
as $A_2(x)=a_2 x^2 +a_1 x +a_0$ and
$A_1(x)=b_1 x +b_0$ to obtain the equation
\begin{eqnarray}\label{eq:Sch1} \nonumber
Y''(u)+Y(u) \left[
\frac{a_2-b_1+2\gamma}{2\omega^2(u)}+ 
\frac{\omega '(u)^2}{4\omega^2(u)} - 
 \frac{\omega ''(u)}{2\omega (u)} \right.  \\ 
-\; 
  {\frac{3\,{{\left( {a_1} + 2\,{a_2}\,F(u) \right) }^2}}
    {16\omega^2(u)\,\left( {a_0} + {a_1}\,F(u) + 
{a_2}\,{{F(u)}^2}
         \right) }}  \\ \nonumber
+ \; 
{\frac{\left( {a_1} + 
2\,{a_2}\,F(u) \right) \,
      \left( {b_0} + {b_1}\,F(u) \right) }{2\omega^2(u)\,
      \left( {a_0} + {a_1}\,F(u) + {a_2}\,{{F(u)}^2} \right) 
      }}  \\ \nonumber
\left. -\; {\frac{{{\left( {b_0} + {b_1}\,F(u) 
\right) }^2}}
    {4\omega^2(u)\,\left( {a_0} + {a_1}\,F(u) + 
{a_2}\,{{F(u)}^2}
        \right) }} \right]=0
\end{eqnarray}

Now, we start to analyze first systematically 
the simplest case putting function $\omega(u)$
to be  unity.
Then, for the potential of the Schr\"odinger 
equation we have
\begin{eqnarray}\label{eq:pot}
 V(u)=-{\frac{a_2}{2}} +
  {\frac{b_1}{2}} +
  {\frac{3\,{{\left( a_1 + 
           2\,a_2\,F(u) \right) }^2}
      }{16\,\left( a_0 + 
        a_1\,F(u) + 
        a_2\,{{F(u)}^2} \right) }} + \;\;\;\\ \nonumber 
   {\frac{\left( a_1 + 
        2\,a_2\,F(u) \right) \,
      \left( b_0 + 
        b_1\,F(u) \right) }{2\,
      \left( a_0 + 
        a_1\,F(u) + 
        a_2\,{{F(u)}^2} \right) }}+
   {\frac{{{\left( b_0 + 
          b_1\,F(u) \right) }^2}}
      {4\,\left( a_0 + 
        a_1\,F(u) + 
        a_2\,{{F(u)}^2} \right) }}.
\end{eqnarray}
We can explicitly find the dependence $x=F(u)$ 
by solving the
equation  (\ref{transf0}).
%
%
%
Then, taking the inverse function, we get for $x$
\begin{eqnarray}\nonumber\label{eq:xu0}
x=F(u)=\left\{\matrix{ 
 {\frac{{a_1}\,{{\sinh ({\frac{{\sqrt{{a_2}}}\,
 u}{2}})}^2} +\sqrt{a_0 a_2}\,\sinh ({\sqrt{{a_2}}}\,u)}
{{a_2}}}, \;\;& a_2\neq 0, \; {\cal D}\ne0 \\ \nonumber
\frac{-a_1 + (a_1 + 2\,a_2) \,
         e^{\sqrt{a_2}\,u}
     }{2\,a_2}, \;\;& a_2\neq 0, \;{\cal D}=0  \\ \nonumber
\sqrt{a_0}\,u +\frac{a_1 u^2}{4}, & a_2=0,\,a_1 \neq 0 
\;\;\;\;\\ 
\frac{1}{\sqrt{a_0}}\, u,  \;\;\;\;\;\; & a_2=0,\, 
a_1=0, \, a_0\neq 0
}\right.
\end{eqnarray}
where ${\cal D} =a_1^2-4a_0a_2$
is the discriminant for $A_2(x)$.

The explicit expression for the quantum potential 
can be obtained after the substitution of 
(\ref{eq:xu0}) into expression (\ref{eq:pot})
but for the general case the resulting formula  becomes 
too complicated.
Specifying the values of the parameters $a_i,b_i$ it is
 possible to show that almost all
known solvable cases in quantum mechanics except the 
Coulomb potential,
are inside the potential family we constructed. The latter
 one will be analyzed later
when we try to construct new potentials by choosing 
a nontrivial $\omega(u)$ function. So let us first investigate the case  
 $\omega(u)=1$.

We shall classify the different cases by the roots 
of the polynomial $A_2(x)$. 
We have two topologically different cases when $A_2(x)$ 
is not degenerate, namely
when $A_2(x)$ has {\it two different roots and the discriminant
${\cal D} \ne 0$}, 
and the case of {\it one root 
with degeneracy 2, ${\cal D} =0$}. 
As one can easily
see in the space of the parameters $a_i$ the first  case
fills the region inside and outside the conical surface for which 
the equation 
is ${\cal D}=0$. So we call the appropriate
cases regular and irregular (for ${\cal D}=0$), respectively. 
We shall refer to them sometimes
as  Jacobi and Morse cases, based on the name of the 
appropriate polynomials (for
the first one) and solutions (second one). 
 
The additional cases appear in (\ref{eq:xu0}) as a result of the degeneracy 
of the polynomial $A_2(x)$, 
it could be of the first order
(we call this case Laguerre's case; $a_2=0$) and  of the zero order 
(further referred to as Hermite case; $a_2=a_1=0$).

The first remark we would like to make is that, our polynomial 
family has five parameters
whereas spectra depend on top power coefficients of $A_2(x), 
A_1(x)$ only,
namely $a_2,b_1$ (see equation (\ref{eq:gamma})), and so  we have the
evident
freedom of choosing  some  parameters without loosing characteristic 
features of the
problem. Obviously, we can change parameters $b_0,a_1$ simply by the
trivial change of the
origin in $x$ variable and the scale on it. Then, if we choose 
definite values of $a_2,b_1$
(one of them could be considered as a scale for energy and could 
be chosen e.g. as unity)
we obtain a two parametric family of polynomials and one parametric family
with full 
isospectrality property. The variation of the
parameter $a_0$ then will lead to different non-trivial 
cases we mentioned.

So, we start with the regular case  ${\cal D}\ne 0$. 
Then the change of variables $x\to u$ is given by the top 
line in (\ref{eq:xu0}), and the
orthogonal polynomials have the weight function given by
\begin{eqnarray}\label{eq:weight1}
 W(x)= \frac{\exp \left\{ 
  \frac{ \left(2\,a_2\,b_0 - a_1\,b_1 \right) }{a_2\,{\cal D}}
   \arctan(\frac{a_1 + 2a_2 x}{\cal D})
  \right\}}{
 {( a_0 + a_1\,x + a_2\,x^2)}^{1-\frac{b_1}{2a_2}}}
\end{eqnarray}
As this family of polynomials has no commonly used name we will refer to 
it as to
generalized Jacobi polynomials, ordinary Jacobi case corresponds to 
$a_0=1, a_1=0,a_2=-1$,$b_0+b_1=2p,b_0-b_1=2q$ corresponding 
to symmetrically chosen
real roots at $\pm 1$ and the interval of orthogonality $[-1,1]$.

The quantum potential has the following general form 
($z=\sqrt{a_2}u$)
\begin{eqnarray}\label{eq:pot1}
 V(u)=
 \frac{A + B \sinh z + C \cosh z + 
           D \sinh 2z + E \cosh 2z }{
      \left( 2\sqrt{a_0a_2}\, \cosh z + a_1\,\sinh z  )
            \right) ^2} 
\end{eqnarray}
with the coefficients $A,B,C,D,E$ expressed through the 
original ones as
\begin{eqnarray} \nonumber
A =
&{{{a_2}}^2}\,\left( 5\,{{{a_1}}^2} - 20\,{a_0}\,{a_2} + 
     8\,{{{b_0}}^2} \right) 
+ 
  \left( 3\,{{{a_1}}^2} - 4\,{a_0}\,{a_2} \right) \, 
   {{{b_1}}^2}\,  \\   \nonumber
 &+\;2\,{a_2}\,\left( -3\,{{{a_1}}^2} + 
  12\,{a_0}\,{a_2} - 
     4\,{a_1}\,{b_0} \right) \,{b_1} , \;\;\;\;\;\;\;
\;\;\;\;\;\;\;\;\;\;\;\;\;\;\;  \\ \nonumber
B =&
8\,{\sqrt{{a_0}\,{a_2}}}\,
  \,\left({b_1}- 2\,{a_2} \right) \,
     \left( 2\,{a_2}\,{b_0} - {a_1}\,{b_1} \right) , \;\;\;
\;\;\;\;\;\;\;\;\;\;\;\;\;\;\;\;\;\;\;
    \\
C =&-4\,{a_1}\,\left( 2\,{a_2} - {b_1} \right) 
      \,\left( 2\,{a_2}\,{b_0} - 
          {a_1}\,{b_1} \right) ,  \;\;\;\;\;\;\;\;\;\;\;
\;\;\;\;\;\;\;\;\;\;\;\;\; \\\nonumber
D =&
4a_1\,{\sqrt{{a_0}\,{a_2}}}\,
   \,{{\left( {a_2} - {b_1} \right) }^2},  \;\;\;\;\;\;
\;\;\;\;\;\;\;\;\;\;\;\;\;\;\;\;
 \;\;\;\;\;\;\;\;\;\;\;\;\;\;\;\;\;\;\;\;\;\;\\\nonumber
E =&\left( {{{a_1}}^2} + 4\,{a_0}\,{a_2} \right) 
        \,{{\left( {a_2} - {b_1} \right) }^2}.  \;\;\;
\;\;\;\;\;\;\;\;\;\;\;\;\;\;
\;\;\;\;\;\;\;\;\;\;\;\;\;\;\;\;\;\;\;\;\;\;\;
\end{eqnarray}
It is straightforward to see that  the potential family 
(\ref{eq:pot1}) includes 
P\"oschl-Teller potentials (both ordinary and modified), 
Scarf-like potentials,
Rosen-Morse and Manning-Rosen potentials \cite{Grosche-Steiner98} at 
appropriate choice of the  parameters.

In singular case $D=0$ it is more convenient to introduce 
other parameters 
rather than $a_i$, 
namely $a_0=\alpha^2, a_2=\beta^2, a_1=2\alpha\beta$, automatically 
satisfying the degeneracy
condition, and then the weight function becomes
\begin{eqnarray}\label{eq:weight2}
W(x)=\left(\alpha+\beta x\right)^{-2+b_1}\exp \left\{\frac{b_1 
\alpha-b_0 \beta}{\beta^2(\alpha+\beta x)} \right\},
\end{eqnarray}
and the potential reads through newly introduced coefficient 
$A,B,C$
\begin{eqnarray}\label{eq:pot3} \nonumber
 V(u)=A + Be^{-\beta  u} + C e^{- 2\beta u}, \\\nonumber
 A=\frac{\left(b_1 - \beta^2 \right)^2}{4\,\beta ^2}, \\ 
%
 B= -
\frac{ \left(b_1\,\alpha  - b_0\,\beta  \right) \,
       \left( b_1 - 2\,\beta ^2 \right) }{2\,\alpha \,\beta^2} \\\nonumber
%
 C=
\frac{\left( b_1\,\alpha  - b_0\,\beta  \right)^2}{4\,\alpha ^2\,\beta^2}. 
\end{eqnarray}
This evidently corresponds to the Morse  class of potentials 
\cite{Landau3}.

The case when $A_2(x)$ becomes  the first order polynomial ($a_2=0$), 
gives for the weight the
following formula
\begin{eqnarray}\label{eq:weight3}
W(x)= e^{\frac{b_1\,x}{a_1}}\,
\left( a_0 + a_1\,x \right)^{-1+\frac{b_0}{a_1}-
\frac{ a_0\,b_1}{a_1^2}},
\end{eqnarray}
and for the potential
\begin{eqnarray}\label{eq:pot3a}
V(u)= A+ B u^{-2} +C u^{2}, \;\;\;\; \;\; \;\; \;\;
\end{eqnarray}
with the new parameters $A,B,C$ (not to be confused with those
obtained above) expressed through the old ones as
\begin{eqnarray}\label{eq:par3} \nonumber  \matrix{
A&= \;\;\;\frac{b_1(a_1b_0 - a_0b_1)}{2a_1^2}, \;\;\;\;\;\;\;\;\; 
\;\; \;\; \;\;\;\;\;\; \;\;  
\\  \nonumber
B&= \;\;\frac{(a_1^2-2a_1b_0+2a_0b_1)(3a_1^2-2a_1b_0+2a_0b_1)}{4a_1^4},  
\\ 
C&= \;\; \;\;\frac{b_1^2}{16}. \;\;\;\;\;\;\;\; \;\; \;\; 
\;\; \;\; \;\;
 \;\; \;\;\;\;\;\; \;\; \;\; \;\
}
\end{eqnarray}

The resulting potentials, as we see, are the combination of 
harmonic oscillator (HO) potential plus
centrifugal potential $B/u^2$. 
And for the sake of completeness it is worthwhile to mention, 
that the case when $A_2(x)$ is a constant ($a_2=a_1=0$)
corresponds to  ordinary HO case with the oscillator position 
shifted by  $-b_0/b_1$.

Before going further, let us construct the explicit representation 
for the wave functions
of the appropriate Schr\"odinger equation and let us discuss the bound 
states within this approach.
As we made two subsequent transformations to obtain the 
Schr\"odinger equation,
the solution in terms of polynomials has the form
\begin{eqnarray}\label{eq:gen-sol}
Y(u)=\exp\left\{ \chi(u)\right\} P_n(F(u))
\end{eqnarray}
where $\chi(u)$ is given by (\ref{transf1}) and $F(u)$ is 
given by (\ref{transf0}). 
The energy corresponding to
this eigenfunction turns out to be the sum of $\gamma_n$ 
( see equation (\ref{eq:gamma}) )
and some constant
factor depending upon the parameters $a_i, b_i$ and leading to the shift 
of the energy's origin.
As the family of orthogonal polynomials has infinity and 
countable number of members,
though the class of the potentials includes not only 
those which grow indefinitely at infinity (supporting bound states only),
but also such with a
finite number of levels (and finite ionization energy), 
we have to understand what is the condition for a bound state 
in the system.
Indeed, this is very simple in the discussed case, the function 
$W(u)$  gives
the asymptotic behaviour of the ground state wave function and 
as the point transformation
$x=F(u)$ could be  non-trivial, the resulting high order 
polynomials $P_n(F(u))$ can have
growing behaviour at infinity which might more than compensate that of
$W(u)$ 
and thus makes $Y(u)$'s
norm infinite. Therefore, the condition for the bound states 
is simply the condition
of a finite norm of $Y(u)$.
Of course, an interesting question appears whether
the non-normalizable solutions of polynomial type correspond 
to some physically significant
features of the system, e.g. to quasi-bound states 
(long-lived localized states) embedded in the continuum, 
but we will not discuss
it here.

Now, we can consider other possible choices of the 
function $\omega(u)$. This is stimulated by the known
sequence of transformations for the Coulomb problem \cite{Landau3}, 
where the first step is the change of scale  in a way
to put "energy" parameter into potential function with 
subsequent transformation of the original
Schr\"odinger equation into hypergeometric equation.

As we can see from the equation (\ref{eq:Sch1}) the problem is 
pure algebraic and
there are several ways to try to obtain the free parameter 
which could be interpreted as
energy.
The first one is to choose $\omega(u)$ to satisfy the equations 
either $\omega'(u)/\omega(u)=k$ or
$\omega''(u)/\omega(u)=k$, that will produce constant factor due to 
fractions including the appropriate
ratios in equation (\ref{eq:Sch1}).
The second one takes place when 
$\omega^2(u)=A_2(F(u))^{-k}$,  $k>0$,
that could lead to free coefficient in potential due to cancellation 
of some denominators in
(\ref{eq:Sch1}). 
We shall not pursue  these cases any further, but treat the most 
important case instead.
Namely, if the order of polynomial $A_2(x)$ is less than two,
then  new possible cases also appear, as we will see, for $k=-1$, which
turns out
to be precisely  the Coulomb case that we discuss below.
The last and more special case is realized 
when $A_2(x)$ has different real roots and the coefficients
$b_0,b_1$ are chosen in such a way as to construct common 
divisor (of the first order) 
for both numerators and denominators in potential. 
In this case choosing $\omega(u)$ in the form of
$\omega(u)=(F(u)-x_1)$ we also obtain free coefficient 
in potential. We will not consider all the above mentioned
cases in detail here but restrict ourselves to one specific choice of
$A_2(x)$, stimulated by the Coulomb problem.

Let us assume that $A_2(x)=x$ and we will choose the
$\omega (u)$ in the form $\omega^2 (u)=F(u)^{-k}$. 
Then, if we choose $k=-1$, it is easy to see from the equation
(\ref{transf0})
that the point canonical transformation  turns out to be identity,
and we get the standard Coulomb case
\begin{eqnarray}
 Y''(u)+\left[-\frac{b_1^2}{4} + \frac{b_0}{2u^2} - 
\frac{b_0^2}{4u^2} - 
    \frac{b_0 b_1}{2u} + \frac{\gamma}{u}\right]Y(u)=0.
\end{eqnarray}

The case $k=1$ is also a special case here, so we have
the following expression for the variable change
$x= e^{-u}$ and the Schr\"odinger equation takes the form
\begin{eqnarray}
 Y''(u)+\left[
-\frac{({b_0}-1 )^2 }{4} - 
  \frac{\left( {b_0}\,{b_1} - 2\,\gamma  \right)}{2}\; e^{-u} - 
  \frac{b_1^2 }{4}\; e^{-2\,u},
\right]Y(u)=0
\end{eqnarray}
which is the case of the quantum Morse potential.
 
For  $k\ne \pm 1$ we obtain after integration of the equation
(\ref{transf0})
the following expression for the point transformation,
\begin{eqnarray}\label{eq:trL}
 x=\left[\frac{(k-1)( u - C_1)}{2}\right]^{\frac{2}{1 - k}}.
\end{eqnarray}
The substitution of the last expression leads to a fairly 
complicated form of  the Schr\"odinger 
equation for $Y(u)$, but as one can show, 
there are no more cases except those we mentioned, where it is possible to
obtain a 
free parameter in 
the role of "energy".

The successful implementation of the construction of solvable 
Schr\"odinger potentials,
as we already have seen, was due to the evident existence of 
polynomial solution of the
equation (\ref{eq:Fi}). 
It is possible to find the generalization  in more complicated 
cases, which is the topic of the following section.

\section{Some generalizations of the approach}
As we saw, the main feature of the system considered 
was that 
the second order differential operator $\hat {\cal L}$ 
(see equations (\ref{eq:SA-LO1}) and (\ref{eq:Fi}))
preserves the linear subspace ${\cal M}_n$ of the polynomials 
of order $n$ for all $n$. It was due to the special adjustment of 
the orders of polynomial
coefficients with the order of appropriate  derivatives. 
This idea, of course, can be
implemented
not only for a special case of SODE and polynomial coefficients 
up to the second order, but in
much more general case of linear PDEs. 
Indeed, we can construct the following general form of
the $n$-th order linear differential operator $\hat {\cal L}$ with 
the same property,
so that it preserves the space when 
acting in the space of the polynomials of
$m$ variables ${\vec x}=\{x_1,...x_m\}$. The general form of 
the appropriate linear PDEs
reads 
\begin{eqnarray}\label{eq:PDE0}
 \hat {\cal L} \, Y({\vec x})\; =\;\sum_{j=0}^{n} P_{j+N}
({\vec x})\partial_{\kappa_j} Y({\vec  x}
\,) =\, 0,
\end{eqnarray}
where 
$N$ is some non-positive integer number 
(when $N<0$ we have degenerate cases, see analogous discussion on SODE in
the      
previous section).
We introduce multi-index $\kappa_j$ as $j$-th order partial 
derivative
over arbitrary combinations of variables in a standard  way, 
by
\begin{eqnarray}
\partial_{\kappa_j}=\frac{\partial^{i_1}}{\partial x_1^{i_1}} 
\cdots
                    \frac{\partial^{i_m}}{\partial x_m^{i_m}}, 
\;\;\;\;\;\;\;\;\;\;\;\;
 j=i_1+\cdots i_m.
\end{eqnarray}

We also  say that the weight of $\kappa_j$ equals $j$ and write 
it as $ \#\kappa_j=j$.
It is worthwhile to point out that we even do not need to demand 
the commutativity of
the derivatives, so that the same consideration could be applied  
for non-commutative case (quantum groups and quantum algebras 
see, e.g. 
\cite{Azcarraga-Izquierdo95}). 
Moreover, we can consider the operators which do not preserve 
such finite dimensional
spaces, but map one into another (with higher dimension). 
The latter is just the permission for
$N$ to be positive.

In the case of standard partial derivatives, when we are 
looking for
the solution in terms of  $k$-th order polynomials in the 
ring $F[x_1,x_2,..,x_m]$,
every term in the sum in (\ref{eq:PDE0})
maps the argument  into the space spanned by the monomials 
$x^{\kappa_{N+k}}=x_1^{i_1}x_2^{i_2}...{x_m}^{i_m}, \;\; 
\{i_1+\cdots i_m=N+k\}$.
The latter space is a  finite dimensional vector space and a
direct sum of the spaces of 
symmetric homogeneous polynomials corresponding to different permutations of
indices for
the monomials  written above.
We denote the space spanned by the definite monomials  of order $k$ as
 $^{(k)}{\cal T}_{[i_1\dots i_m]}$).
Then we can write down the
expression for the dimension for the image space for operator action 
\begin{eqnarray}
 dim \;{\cal M}^N_k = \sum\limits_{j=0}^{k+N} 
 \sum\limits_{{\stackrel{i_1,\cdots,i_m=0}{ i_1+ \cdots +i_m=m}}}^{m}\;
dim\; 
^{(j)}{\cal T}_{{[i_1\dots i_m]}}.
\end{eqnarray}

Now, the construction of the k-th order polynomial solution 
\begin{eqnarray}
Y_k({\vec x})= \sum\limits_{j=0}^{k} C_{\kappa_j} 
x^{\kappa_j}
\end{eqnarray}
leads simply to the linear algebraic problem for non-trivial 
solution for the
coefficients $C_{\kappa_j}$. 

At this point two different cases are possible.
The first one is realized if $N \le 0$, that is the maximal 
order of polynomial coefficient
is less or equal to the PDE's order. 
In this case we can always satisfy the system of equations
because the number of linear homogeneous
equations for $C_{\kappa_j}$
%
%
is precisely equal to $ dim\; {\cal M}^0_k$.
Then, the non-triviality condition is the condition of zero 
determinant for corresponding
matrix obtained from equating all coefficients at monomials 
of type $x^{\kappa_i}$
to zero, and this gives us spectral parameter for the 
polynomial family,
namely the quantization condition imposed on 
the coefficient $P_0(x)$. Then, considering the problem
over the field of complex numbers,
 we can always  construct a polynomial family in this case
for some quantized value of the coefficient $P_0(x)$.
In the contrary, when we have the condition  $N > 0$, we are 
still obliged 
to fulfill $ dim\; {\cal M}^N_k$
conditions but  only  for $ dim\;{\cal M}^0_{k}$ 
coefficients $C_{\kappa_j}$.
The resulting system becomes overcomplete which simply means that 
{\it we can construct
some separate polynomial solutions} of the equation 
(\ref{eq:PDE0}) for only 
{\it  a few levels, maybe even  one},
that is for definite choice of  $n$ and, additionally 
for special values of some 
of the  
coefficients in the coefficient polynomials.

It is very interesting to mention here that in the 1-D case considered 
in the previous section for SODE, the  appropriate matrix
turns out to be upper-three-diagonal, with additional
relationship between elements, so that its determinant for $j$-th order
system
has the following form (preserving notions for coefficient of
$A_2(x),A_1(x)$)
\begin{eqnarray}\nonumber
\left(
 \matrix{
    &\gamma  &b_0     \;\;\ \;\; \;\;\;\       2 a_0  &\dots   & 0
\\ \nonumber
    &0       &\gamma+ b_1  \;\;\   2 b_0      &\dots   &0          \\
\nonumber
    &\dots &\dots &\dots &\dots    \\ \nonumber
    &0       &\dots        &(j-1)b_0+(j-1)(j-2)a_1          &j(j-1)a_0 \\
\nonumber
    &0       &\dots    &\gamma+(j-1)b_1+(j-1)(j-2)a_2   &j b_0 + j(j-1)a_1
\\ \nonumber
    &0       &\dots           &0                               &\gamma+j b_1
+ j(j-1)a_2  
 }
\right)
\end{eqnarray}
that leads to one degenerate eigenvalue 
(rather than $j+1$)  for a given $j$-th order  polynomial.

What realizations of the scheme described above could be successfully 
used for the construction of
the solutions for Schr\"odinger equation except those which we 
demonstrated in Section 2?
There are three evident but not  easy ways. 
The first and the easiest one is the consideration
of 1D problems with higher order polynomials, and  
we intend to 
demonstrate this in one
example,  putting away its full description as a subject 
of separate
publication. We consider the construction 
of the polynomial solutions
starting from the third order polynomials to give the 
representation of the problems
emerging there. Let us have the equation of the
form
\begin{eqnarray}  \label{eq:3d-order0}
 x^3 y''(x) + \alpha(x^2-1) y(x) +(\beta x + \gamma) y(x)=0. 
\end{eqnarray}
Then, in the same manner as we did  in Section 2 we map the equation 
to the Schr\"odinger equation  via two
subsequent  point transformations  $x=4/u^2$, and  the gauge 
transformation  
\begin{eqnarray}
 \chi (u)= C_1+\frac{\alpha u^4}{64} +
 \frac{3-2\alpha}{2}\log u.
\end{eqnarray}
The resulting equation has the form
\begin{eqnarray} \label{Schr3o}
 Y''(u) +\left( \gamma- V(u)\right)Y(u)=0, \\
 V(u)={\frac{(3\alpha - \alpha^2)u^2 }
    {8}} + {\frac{{u^6}\,{{\alpha }^2}}{256}} + 
  {\frac{ 4\,{{\alpha }^2}- 8\,\alpha - 
16\,\beta +3 }{4\,{u^2}}}.
\end{eqnarray}
The solution will be given by the formula (\ref{eq:gen-sol}), 
but now 
we have to construct  the polynomials in a non-trivial way, 
because the Rodrigue's formula is no more
applicable. So we start to look directly for the polynomial 
solution of the equation (\ref{eq:3d-order0}). 
Let us define $n$-th order polynomial
as $p_n(x)=\sum_{i=0}^{n} c_i x^i$. 
When we substitute this anzatz into equation 
(\ref{eq:3d-order0})
and equate to zero the coefficients at all orders of 
independent variable $x$,
we obtain $n+1$ equations for $n$ coefficients 
$c_i, \,i=0\dots n-1, \; \gamma$ 
(the last coefficient $c_n$ should be chosen in order to satisfy
the standardization condition). 
Therefore the system is overcomplete
and for a non-trivial solution we must 
specify some additional coefficient in a unique way or to impose one
additional
condition on some parameters of the system.
In our case we have the only choice to add $\beta$ to the list 
of coefficients to be found.
Let us find then the solution for example for $n=1,2$ explicitly starting
from the
case $n=1$.

Solving the set of equations for $c_0,\beta,\gamma$, two solutions
$c^{\pm}_0=\pm\, 1, \;\beta=-\alpha, \;
\gamma^{\pm}=\mp\, \alpha$ 
can be constructed that giving for  $Y_1(u)$
\begin{eqnarray} 
Y^{\pm}_1(u)=e^{\frac{\alpha u^4}{64}}
     \left( \mp 1 + {\frac{4}{{u^2}}} \right)
     u^{\frac{3 - 2\,\alpha }{2}}, \;\;\; (\gamma=\mp \alpha) .
\end{eqnarray}
We have to impose the finiteness condition for the norm of the solution, so
that 
in our case $\alpha<-1/4$. The last requirement
follows from simultaneous demand on proper behaviour at infinity 
that leads to $\alpha<0$
and integrability at $u=0$.
But as one can see, in the interval $-1/2 <\alpha\le -1/4$ the potential
becomes
repulsive as $u\to 0$. The last property means that we have to impose the
boundary condition
$Y(u)|_{u=0}=0$ that leads to the  restriction  $\alpha\le -1/2$  
on the admittable region for the variation of $\alpha$.

Then, both solutions will be, as 
one can see, 
bound states of the system and  the 
solution for larger $\gamma$
will have zero not only at $u=0$ but also at $u=2$, which evidently 
corresponds to the first
excited state, whereas the first one corresponds to the ground state.

In a similar way, for $n=2$ we obtain the condition 
$\beta= -2\,\left( 1 + \alpha  \right) $
and the following three solutions 
for  $c_0,c_1,\gamma$
\begin{eqnarray}
 c^{(1)}_0    = -\frac{\alpha }{\alpha+1 }, \;\;\
 c^{(1)}_1    =  0\;\;\;\;\;
 \gamma^{(1)} = 0,\;\;\;\; \;\;\;\;\;\;\;\;\;\;\;\;\;\;\\ \nonumber
 c^{(\pm)}_0  = \frac{\alpha }{ \alpha+2 }, \;
  c^{(\pm)}_1    =
    \pm \frac{\sqrt{2\alpha\left(2\,\alpha +3 \right)}}
            {\alpha  + 2}, \;\;
 \gamma^{(\pm)} = 
    \pm 2\,\sqrt{2\alpha\left(2\alpha +3 \right)} .
\end{eqnarray}

Then, the appropriate eigenfunctions are given by
\begin{eqnarray}
 Y^{(2)}(u)=  {e^{{\frac{\alpha u^4 }{64}}}}\,{u^{{\frac{3}{2}} - \alpha
}}\,
   \left({\frac{16}{{u^4}}} - {\frac{\alpha}{\alpha +1 }} \right) , \\
\nonumber
 Y^{(\pm)}(u)= {e^{{\frac{\alpha{u^4} }{64}}}}\,{u^{{\frac{3}{2}} - \alpha
}}\,
   \left( {\frac{16}{{u^4}}} + {\frac{\alpha }{\alpha +2 }} \pm 
     \frac{4\sqrt{2\alpha\left(2\,\alpha +3 \right)}}{(\alpha+2)u^2}
\right). 
\end{eqnarray}
The similar consideration as we performed
for $n=1$ shows that admittable region for the parameter $\alpha$ now is
given
by $\alpha\le -5/2$. Then the constructed eigenstates represent the ground
and
the first two excited states for the potential
\begin{eqnarray}
V(u)=\frac{\alpha ^2  u^6 }{256}-
     \frac{\alpha\left(\alpha  -3\right){u^2}}{8} +  
  \frac{4 \alpha^2+ 24\,\alpha+35  }{4\,u^2}.
\end{eqnarray}
The remarkable feature of the example we considered was that we construct
some 
eigenstates corresponding to a given potential {\it using polynomial anzatz
of a given order}. The solution for those cases corresponding to the ground
state and to the excited states  was given by the number of the roots of the

polynomials of a given order.
This is in  contradistinction  with the standard solvable situation
where polynomial's order  is equal to the  quantum 
number because for the classical
polynomial family, the $n$-th order polynomial has precisely $n$  roots 
 on an interval where the family  is  defined, unlike  in the general case, 
where the real polynomial can have complex (nonreal) roots.

>From the last we can make some useful conclusions. As we see, if the
order of polynomial coefficient functions is different 
from the order of the equation, the polynomial solutions in general 
do not exist
except for the special values of the parameters. In the example that has 
been considered,
we have to put one additional condition on the parameter to 
construct non-trivial solution.
Nevertheless, after appropriate restriction of the region for some
parameters
($\alpha$ in the considered example) and by fixing the value for some
other (expressing $\beta$ through the $\alpha$ in the discussed case)  
we were able to construct eigenfunctions of some low lying eigenstates

Now, we return to the discussion of other possible generalizations of the
proposed approach.

The second admissible choice is to consider, in the same way as we did,  
the Schr\"odinger
equations solvable in momentum representation. As it is easy 
to understand,  physically
interesting potentials  of polynomial type with the order greater 
than two will
correspond to higher order differential equations, 
so that we can e.g. ask which
Schr\"odinger equations can be constructed, based 
on the polynomial solutions for
the equation like this $A_4(x) y''''(x)+A_3(x)y''' +...=0$? 
We stop the discussion
of this possibility at this point leaving it for
the future publications too.

The last and the most complicated case corresponds 
to the consideration of
genuine higher dimensional problems and appropriate PDEs. 
The most difficult obstacle to be overcome there is
the necessity to perform transformations from original 
equation for polynomials to the
equation of the  Schr\"odinger type. Although the theory of 
characteristics is  applicable
in this case, the resulting equation, at first glance,  
could be hardly interpreted in terms of the
Schr\"odinger type. As for the latter one we must demand 
the existence of pure constant term
included into coefficient function at zero-derivative 
(see the discussion after equation (\ref{eq:Sch0})).
Nevertheless, this direction is of great importance for 
deeper insight into integrability
and solvability problems in quantum mechanics.

\section{Discussion}
 In summary, we demonstrated that one can construct explicit 
formulae for the family of the orthogonal 
 polynomials depending on five parameters, and thus  we can associate 
with them the family of
 isospectral potentials (isospectrality with respect 
to 3 free parameters)
 which include almost all known quantum mechanically exactly solvable
 potentials. Some generalization of the approach to
 higher dimensional equations (PDEs) as well as to the  higher order 
ODEs has been proposed.

We may conclude with a little bit speculative thought.
If one would put forward the requirement for the bound 
states of a quantum system (in analytical case)
 as a demand on polynomial type of the reduction of the wave function 
 (that seems to be reasonable
 and evidently fulfilled for all up-to-date known 1D 
solvable cases),  then
 the immediate conclusion follows that the proposed approach 
(with its generalizations
 discussed) includes {\it all analytically solvable cases}.

\section*{Acknowledgements}
 The authors are grateful to
 the referees for drawing our attention to the papers of Natanzon 
 and Turbiner and for stimulating the improvement of the manuscript.
 This research has been supported by the Ministry of Science 
 and Technology of the Republic of Slovenia. GK acknowledges also the 
postdoctoral research grant of the Slovenian Science Foundation.

\end{document}